\documentclass[12pt]{article}
\usepackage{amsfonts}

\newcommand{\dss}[1]{{\rm{\displaystyle{#1}}}}

\renewcommand{\footnote}[1]{\def\thefootnote{\arabic{footnote})}\footnotemark\footnotetext{#1}}

\newcommand{\Z}{{\Bbb Z}}

\newcommand{\io}{[\hspace{-1pt}[}
\newcommand{\ic}{]\hspace{-1pt}]}
\newcommand{\fo}{\{\!\mid\!}
\newcommand{\fc}{\!\mid\!\}}

\newcommand{\sgn}{{\rm sgn}}

\renewcommand{\Re}{{\rm Re\,}}
\newcommand{\REG}{{\rm REG}}
\newcommand{\IRREG}{{\rm IRREG}}

\newcommand{\ren}{{\rm ren}}
\newcommand{\tr}{{\rm tr}}

\renewcommand{\arctan}{{\rm arctan}}

\renewcommand{\cosh}{{\rm cosh}}

\newcommand{\Lp}{L_{(+)}}
\newcommand{\Lm}{L_{(-)}}

\def\be{\begin{equation}}
\def\ee{\end{equation}}
\def\bea{\begin{eqnarray}}
\def\eea{\end{eqnarray}}
\def\ba{\begin{array}}
\def\ea{\end{array}}
\def\beq{\begin{eqnarray*}}
\def\eeq{\end{eqnarray*}}

\def\I{{\cal I}}
\def\P{{\cal P}}

\def\V{{\bf V}}
\def\j{{\bf j}}

\def\vac{{\rm vac}}
\def\x{{\bf x}}

\def\M{{\cal M}}
\def\N{{\cal N}}
\def\S{{\cal S}}

\def\C{{\cal C}}
\def\J{{\cal J}}

\newcommand{\bpar}{\mbox{\boldmath $\partial$}}

\newcommand{\sss}{\scriptscriptstyle}

\begin{document}

\title{Induced vacuum condensates \\
in the background of a singular magnetic vortex \\
in 2+1-dimensional space-time}
\author{\bf{Yurii A. Sitenko}\thanks{Electronic address: yusitenko@bitp.kiev.ua}\\ 
Bogolyubov Institute for Theoretical Physics,\\
National Academy of Sciences,\\
14-b Metrologichna str., 252143, Kyiv, Ukraine}
\date{}
\maketitle

\begin{abstract}
We show that the vacuum of the quantized massless spinor field in 2+1-dimensional 
space-time is polarized in the presence of a singular magnetic vortex. Depending 
on the choice of the boundary condition at the location of the vortex, either 
chiral symmetry or parity is broken; the formation of the appropriate vacuum 
condensates is comprehensively studied. In addition, we find that current, 
energy and other quantum numbers are induced in the vacuum.\\
PACS numbers: 03.65.Bz, 03.70.+k, 11.10.Kk, 11.30.Rd\\
Keywords: vacuum condensate, chiral symmetry, singular vortex
\end{abstract}

\section{Introduction}

Field models in 2+1-dimensional space-time have been intensively 
explored in recent years. The interest to this subject is motivated 
by its apparent relevance for the description of planar condensed matter 
systems with rather fascinating properties, including that of 
high-temperature superconductivity (see, e.g., Refs.\cite{Sem,Dor,Mack,Far}).  
An important role is played by the study of induced vacuum quantum 
numbers, in particular, induced vacuum condensates exhibiting various 
symmetry breaking patterns. It has been shown that the homogeneous external 
magnetic field strength induces the chiral symmetry breaking condensate in
the vacuum in the universal manner, i.e. irrespectively of all possible 
types of interaction among quantized fermion fields \cite{Gus}. Certain efforts 
have been undertaken in an attempt to generalize this result to the case of 
the inhomogeneous external magnetic field strength \cite{Dun}. The aim of the 
present paper (see also Ref.\cite{Sit9}) is quite different, being inspired by 
the famous Bohm-Aharonov effectm \cite{Aha}  --  we pose the question: whether 
the condensate can emerge in the vacuum even in the case of vanishing external 
magnetic field strength and vanishing interaction among quantized fermion 
fields? The answer will be positive, and we shall show that the vacuum 
condensate can be induced by an external field potential rather than an 
external field strength. Also, all other vacuum polarization effects in the 
background of the Bohm-Aharonov magnetic field configuration will be determined.

The condensate which describes the pairing of massless fermions 
and antifermions in the vacuum is conventionally defined as
\be
\C(x)=i\langle \vac|T\bar{\Psi}(x)\Psi(x)|\vac\rangle,
\ee
where $\Psi(x)$ is the operator of the second-quantized fermion field  
and $T$ denotes the time ordering operation.
In the background of external classical fields, the vacuum
expectation value of the time-ordered product of the fermion
field operators takes form
\be
\langle\vac| T\Psi(x)\bar{\Psi}(y)|\vac\rangle = \langle x|
(\gamma^\mu\nabla_\mu)^{-1} |y \rangle,
\ee
where $\nabla_\mu$ is the covariant derivative in this background.
Thus, in a static background, condensate (1) is presented as
\be
\C(\x)=\tr\langle x|\gamma^0 (i\partial_0 - H)^{-1}| x\rangle,
\ee
where $x=(x^0,\x), \,  \gamma^\mu = (\gamma^0,{\mbox{\boldmath $\gamma$}}), \, 
\nabla_\mu = (\partial_0,{\mbox{\boldmath $\nabla$}})$ and 
\be
H=-i\gamma^0 {\mbox{\boldmath $\gamma\cdot\nabla$}}.
\ee
After performing the Wick rotation of the time axis ($x^0=-i\tau$), 
Eq.(3) is recast into the form 
\be
\C(\x)=-{1\over2} \,\tr\langle\x|\gamma^0\,\sgn(H)|\x\rangle,
\ee
where
\beq
\sgn(u)=\left\{\ba{cc}
1,& u>0\\ -1,&u<0\\ \ea \right\}.
\eeq
Since Hamiltonian $H$ (4) anticommutes with matrix $\gamma^0$, 
\be
[H,\gamma^0]_{+}=0,
\ee
one could anticipate that condensate (5) vanishes. However, this might 
not be the case for some specific background field configurations. 

It is instructive to rewright Eq.(5) as 
\be
\C(\x)= {i\over4}\,{\mbox{\boldmath $\nabla$}}\,\cdot\tr\langle\x|
{\mbox{\boldmath $\gamma$}}\,|H|^{-1}|\x\rangle.
\ee
Although all $\gamma$-matrices are traceless, current
\be
{\mbox{\boldmath $\J$}}(\x)= {i\over4}\,\tr\langle\x|
{\mbox{\boldmath $\gamma$}}\,|H|^{-1}|\x\rangle,
\ee
can be nonvanishing, then its nonconservation results in the 
emergence of vacuum condensate (7). 

A peculiar feature of the 2+1-dimensional quantum field theory consists 
in a possibility to define chiral invariant vacuum condensate 
(see, e.g., Ref.\cite{Appe})
\be
\P(x)={i\over2}\,\langle \vac|T\bar{\Psi}(x)[\gamma^3,\gamma^5]_{-}\,
\Psi(x)|\vac\rangle,
\ee
which, in a static background, is reduced to the form
\be
\P(\x) = -{1\over4} \,\tr\langle\x|\,\gamma^0\,[\gamma^3,\gamma^5]_{-}\,
\,\sgn(H)|\x\rangle = {\mbox{\boldmath $\nabla$}}\cdot{\mbox{\boldmath $\I$}}(\x),
\ee
where
\be
{\mbox{\boldmath $\I$}}(\x) = {i\over8}\,\tr\langle\x|\,
{\mbox{\boldmath $\gamma$}}\,[\gamma^3,\gamma^5]_{-}\,|H|^{-1}|\x\rangle.
\ee
Here, $\gamma^3$ is the $\gamma$-matrix corresponding to the missing ($x^3$) 
dimension and
\be
\gamma^5 = i \gamma^0 \gamma^1 \gamma^2 \gamma^3.
\ee
This vacuum condensate is directly related to the vacuum spin and breaks parity.

In the present paper we
consider classical static magnetic field in flat 2+1-dimensional 
space-time, as an external background. Thus, the covariant derivative takes 
form 
\be
{\mbox{\boldmath $\nabla$}}=\bpar-i\V(\x),
\ee
where \V(\x) is the vector potential of the magnetic field. The magnetic 
field configuration is chosen  to be that of a singular vortex placed at 
the origin of twodimensional space (the Bohm-Aharonov configuration):
\be
V^1(\x)=-\Phi^{(0)} {x^2\over (x^1)^2+(x^2)^2}, \quad V^2(\x)=\Phi^{(0)} 
{x^1\over (x^1)^2+(x^2)^2},
\ee
\be
\bpar\times\V(\x)=2\pi\Phi^{(0)} \delta(\x),
\ee
where $\Phi^{(0)}$ is the vortex flux in $2\pi$  units, i.e.
in the London ($2\pi \hbar c e^{-1}$) units, since we use
conventional units $\hbar=c=1$ and coupling constant $e$ is included into
vector potential $\V(\x)$. Evidently, vector potential (14) is undefined at 
the origin, i.e. the limiting value
$\lim_{|\x|\rightarrow 0}\V(\x)$ does not exist. Therefore, point $\x=0$ is 
excluded, and a certain boundary condition has to be imposed at this puncture. 
Note that topology of the punctured plane is characterized by winding
number: $\pi_1=\Z$ (here $\pi_1$ is the first homotopy group and
$\Z$ is the set of integer numbers).

Hamiltonian (4) in the background of singular magnetic vortex 
(14)-(15) takes form
\be
H =-i\gamma^0\gamma_{r}\partial_{r} -
ir^{-1}\gamma^0\gamma_{\varphi}(\partial_{\varphi}-i\Phi^{(0)}) ,
\ee
where
\begin{equation}
\gamma_{r} = \gamma^{1}\cos\varphi + \gamma^{2}\sin\varphi,
\quad\gamma_{\varphi} = -\gamma^{1}\sin\varphi + \gamma^{2}\cos\varphi ,
\end{equation}
and polar coordinates,
\beq
r = \sqrt{(x^1)^2+(x^2)^2},\quad
\varphi = \arctan(x^2/x^1),
\eeq
are introduced. 

It is natural to require that the Hamiltonian be self-adjoint
operator. Usually, Hamiltonians in singular backgrounds are not essentially 
self-adjoint, and a possibility of their self-adjoint extension (see, e.g., 
Refs.\cite{Akhie,Alb}) has to be explored. If a solution to this problem is 
found, then it yields the most general boundary condition at the 
puncture\footnote{The cases of various nonrelativistic Hamiltonians in 
singular backgrounds are reviewed extensively in monograph
\cite{Alb}. The cases of the massive and massless relativistic threedimensional 
Hamiltonians in the background of a singular magnetic monopole are considered 
in Refs.\cite{Gol,Cal,Gro,Yam}. The case of the massive relativistic 
twodimensional Hamiltonian in background (14)-(15) is considered in Refs.
\cite{Ger,Sit96,Sit97,Sit99}.}.
In the case of Hamiltonian (16), further restrictions are imposed by either 
parity or chiral symmetry conservation. Having specified the boundary 
condition at the puncture, one can find all vacuum polarization effects 
in background (14)-(15).

In the next section we obtain two one-parameter families of boundary conditions 
at the puncture: one is chiral invariant allowing for parity breaking, and 
another is parity invariant allowing for chiral symmetry breaking. Chiral 
symmetry breaking effects are considered in Section 3. Parity breaking effects 
are considered in Section 4. The absence of the twodimensional anomaly is 
demonstrated in Section 5. Results are summarized in Section 6. Some details 
in the derivation of the results are outlined in Appendices A and B.

\section{Boundary condition at the location of a vortex}

In 2+1-dimensional space-time, the Clifford algebra
has two inequivalent irreducible representations which can be differed
in the following way:
\be
i\gamma^0\gamma^1\gamma^2=s, \qquad s=\pm1.
\ee
Choosing matrix $\gamma^0$ to be diagonal 
\be
\gamma^0=\sigma_3, 
\ee
one gets the most general form
\be
\gamma^1=e^{{i\over2}\sigma_3\chi}i\sigma_1e^{-{i\over2}\sigma_3\chi},
\quad \gamma^2=
e^{{i\over2}\sigma_3\chi}is\sigma_2e^{-{i\over2}\sigma_3\chi},
\ee
where $\sigma_1,\sigma_2$ and $\sigma_3$ are the Pauli matrices, 
and $\chi$ is the parameter varying in the interval
$0\leq\chi<2\pi$ to go over to the equivalent representations.
Since the algebra of the Pauli matrices is complete, there is no 
any other $2\times 2$ matrix which 
anticommutes with the above $\gamma$-matrices. Therefore, in order to 
generate chiral symmetry transformation, one has to consider a reducible 
$4\times 4$ representation composed as a direct sum of two 
inequivalent irreducible $2\times 2$ ones (see, e.g., Ref.\cite{Appe}). 
Taking into account Eqs.(19) and (20), we get the most general form 
of the $4\times 4$ $\gamma$-matrices: 
\bea
\gamma^0 = \left(
\begin{array}{c|c}
\begin{array}{cc}
1 & 0 \\
0 & -1 
\end{array}
& {\dss{O}} \\ \cline{1-2}
{\dss{O}} & 
\begin{array}{cc}
-1 & 0 \\
0 & 1 
\end{array}
\end{array}
\right),
\gamma^1 = i \,\left(
\begin{array}{c|c}
\begin{array}{cc}
0 & e^{i\chi_+} \\
e^{-i\chi_+} & 0 
\end{array}
& {\dss{O}} \\ \cline{1-2}
{\dss{O}} & 
\begin{array}{cc}
0 & -e^{i\chi_-} \\
-e^{-i\chi_-} & 0 
\end{array}
\end{array}
\right) ,
\nonumber
\eea
\be
\gamma^2 = s\,\left(
\begin{array}{c|c}
\begin{array}{cc}
0 & e^{i\chi_+}\\
-e^{-i\chi_+} & 0 
\end{array}
& {\dss{O}} \\ \cline{1-2}
{\dss{O}} & 
\begin{array}{cc}
0 & -e^{i\chi_-} \\
e^{-i\chi_-} & 0 
\end{array}
\end{array}
\right),
\ee
where $0\leq\chi_{\pm}<2\pi$. The algebra is completed by adding
\bea
\gamma^3 = i \,s\,\left(
\begin{array}{c|c}
{\dss{O}}& 
\begin{array}{cc}
e^{{i\over2}(\chi_{+} - \chi_{-})} & 0 \\
0 & e^{-{i\over2}(\chi_{+} - \chi_{-})} 
\end{array} \\ \cline{1-2}
\begin{array}{cc}
e^{-{i\over2}(\chi_{+} - \chi_{-})} & 0 \\
0 & e^{{i\over2}(\chi_{+} - \chi_{-})} 
\end{array}
& {\dss{O}}
\end{array}
\right),
\nonumber
\eea
\be
\gamma^5 = i \,\left(
\begin{array}{c|c}
{\dss{O}}& 
\begin{array}{cc}
e^{{i\over2}(\chi_{+} - \chi_{-})} & 0 \\
0 & e^{-{i\over2}(\chi_{+} - \chi_{-})} 
\end{array} \\ \cline{1-2}
\begin{array}{cc}
-e^{-{i\over2}(\chi_{+} - \chi_{-})} & 0 \\
0 & -e^{{i\over2}(\chi_{+} - \chi_{-})} 
\end{array}
& {\dss{O}}
\end{array}
\right);
\ee
note that the representation which is mostly used (see Ref.\cite{Appe}) 
corresponds to $s=1$ and $\chi_{+}=\chi_{-}=0$. Note also that
\be
{1\over2}[\gamma^3, \gamma^5]_{-} = s\,\left(
\begin{array}{c|c}
\begin{array}{cc}
1 & 0 \\
0 & 1 
\end{array}
& {\dss{O}} \\ \cline{1-2}
{\dss{O}} & 
\begin{array}{cc}
-1 & 0 \\
0 & -1 
\end{array}
\end{array}
\right)
\ee
is sometimes called as the $\tau_3$-matrix (see, e.g., Ref.\cite{Dor}). 

One can define the parity transformation 
\be
\Psi(x^0,x^1,x^2) \rightarrow i\gamma^2 \gamma^3\Psi(x^0,x^1,-x^2)
\ee
and the chiral symmetry transformation
\be
\Psi(x^0,x^1,x^2) \rightarrow e^{i\omega\gamma^5} \Psi(x^0,x^1,x^2),
\ee
as well as transformations with matrix $\gamma^5$ replaced 
by $i\gamma^3$, ${1\over2}[\gamma^3, \gamma^5]_{-}$ and the unity matrix.

Using Eq.(21), Hamiltonian (16) is recast into the form
\be
H=\left(\ba{cc}
H_+& 0\\
0& H_-\\ \ea \right),
\ee
where
\be
H_\pm=\left(\ba{cc}
0& e^{i\chi_\pm -is\varphi}[\partial_r-r^{-1}(i\partial_\varphi+\Phi^{(0)})]\\
e^{-i\chi_\pm +is\varphi}[-\partial_r-r^{-1}(i\partial_\varphi+\Phi^{(0)})]&0\\ \ea
\right).
\ee
Equation of motion
\be
(i\partial_0 - H)\Psi(x)=0
\ee
is invariant under the chiral symmetry transformation (as well as under 
transformations generated by $i\gamma^3$ and ${1\over2}[\gamma^3,\gamma^5]_{-}$). 
A single-valued solution to Eq.(28) is presented conventionally as 
\be
\Psi(x)=\sum_{n\in\Z}\int\limits_0^\infty dE E\, e^{-iE x^0}<\x|E,n>a_{En}  
+\sum_{n\in\Z}\int\limits_0^{-\infty} dE E\,e^{-iE x^0}<\x|E,n>b_{En}^+,
\ee
where $a_{En}^+$ and $a_{En}$ ($b_{En}^+$ and $b_{En}$) are
the fermion (antifermion) creation and annihilation operators
satisfying anticommutation relations
\be
[a_{En},a_{E'n'}^+]_+=[b_{En},b_{E'n'}^+]_+={\delta(E-E')\over\sqrt{|EE'|}}
\delta_{nn'},
\ee
and
\be
\langle\x|E,n\rangle=
\left(\ba{c}
f_n^+(r,E)e^{in\varphi}\\
g_n^+(r,E)e^{i(n+s)\varphi}\\
f_n^-(r,E)e^{in\varphi}\\
g_n^-(r,E)e^{i(n+s)\varphi}\\ \ea
\right);
\ee
radial functions $f_n^{\pm}$ and $g_n^{\pm}$ satisfy the system of equations
\bea
e^{-i\chi_\pm}[-\partial_r+s(n-\Phi^{(0)})r^{-1}]f_n(r,E)==E g_n(r,E),
\nonumber
\eea
\be
e^{i\chi_\pm}[\partial_r+s(n-\Phi^{(0)}+s)r^{-1}]g_n(r,E)=E f_n(r,E).
\ee

Decomposing the value of the vortex flux into the integer and
fractional parts,
\be
\Phi^{(0)}=\io\Phi^{(0)}\ic+\fo \Phi^{(0)}\fc, \quad 0\leq\fo\Phi^{(0)}\fc<1
\ee
($\io u\ic$ denotes the integer part of quantity $u$), one can note
that the case of $\fo \Phi^{(0)}\fc=0$ is equivalent to the case of trivial
topology, i.e. absence of the vortex $(\Phi^{(0)}=0)$.\footnote{This confirms
once more the general fact that a singular magnetic vortex is
physically unobservbale at integer values of the vortex flux
\cite{Aha}. It was as far back as 1931 that Dirac used actually this
fact to obtain his remarkable condition for the magnetic monopole
quantization \cite{Dir2}.} In the case of $0<\fo \Phi^{(0)}\fc<1$ the 
condition of regularity at the puncture $r=0$ can be imposed on the modes
with $n\neq  n_0$ only, where
\be
n_0=\io\Phi^{(0)}\ic+{1\over2}-{1\over2}s;
\ee
consequently, we get 
\be
\left(\ba{c} f_n^\pm \\ g_n^\pm \\ \ea \right) ={1\over2\sqrt{\pi}}
\left(\ba{c}
J_{l-F}(kr)e^{i\chi_\pm}\\[0.2cm]
\sgn(E)J_{l+1-F}(kr)\\ \ea \right), \qquad l=s(n-n_0)>0,
\ee 

\be
\left(\ba{c} f_n^\pm \\ g_n^\pm \\ \ea \right)
={1\over2\sqrt{\pi}}
\left(\ba{c}
J_{l'+F}(kr)e^{i\chi_\pm}\\[0.2cm]
-\sgn(E)J_{l'-1+F}(kr) \ea \right), \qquad l'=s(n_0-n)>0;
\ee 
here $k=|E|$, $J_{\rho}(u)$ is the Bessel function of order
$\rho$, and
\be
F=s\fo\Phi^{(0)}\fc+{1\over2}-{1\over2}s.
\ee

Thus, the partial Hamiltonians corresponding to $n\neq n_0$ are 
essentially self-adjoint when defined on the domain of regular 
functions, and that is only the partial Hamiltonian corresponding to 
$n=n_0$ that needs a self-adjoint extension. The method of self-adjoint 
extensions (see, e.g., Refs.\cite{Akhie,Alb}) is employed in Appendix A, 
resulting in a boundary condition which entails the irregular behaviour 
of the mode with $n=n_0$ at the puncture $r=0$. Actually, there 
are two different possibilities to choose a physically reasonable boundary 
condition: one is chiral invariant and parity violating, while another is 
parity invariant and chiral symmetry violating. The chiral invariant choice is
\be
\cos\bigl(s{\Theta\over2}+{\pi\over4}\bigr)\lim_{r\rightarrow 0}(\mu
r)^Ff^\pm_{n_0}=-e^{i\chi_{\pm}}\sin\bigl(s{\Theta\over2}+{\pi\over4}\bigr)
\lim_{r\rightarrow 0}(\mu r)^{1-F}g^\pm_{n_0}, 
\ee
then the irregular mode takes form
\bea
\left(\ba{c} f^\pm_{n_0}\\ g^\pm_{n_0} \\ \ea \right)
={1\over 2\sqrt{\pi[1+\sin(2\nu_E)\cos(F\pi)]}} \, \times
\nonumber
\eea
\be
\times \left(\ba{c}
[\sin(\nu_E)J_{-F}(kr)+\cos(\nu_E)J_F(kr)]e^{i\chi_\pm}\\[0.2cm]
\sgn(E)[\sin(\nu_E)J_{1-F}(kr)-\cos(\nu_E)J_{-1+F}(kr)]\\ \ea \right).
\ee
The parity invariant choice is 
\be
\cos\bigl(s{\Theta\over2}+{\pi\over4}\bigr)\lim_{r\rightarrow 0}(\mu
r)^Ff^\pm_{n_0}=\mp e^{i\chi_{\pm}}\sin\bigl(s{\Theta\over2}+{\pi\over4}\bigr)
\lim_{r\rightarrow 0}(\mu r)^{1-F}g^\pm_{n_0}, 
\ee
then the irregular mode takes form
\bea
\left(\ba{c} f^\pm_{n_0}\\ g^\pm_{n_0} \\ \ea \right)
={1\over 2\sqrt{\pi[1 \pm \sin(2\nu_E)\cos(F\pi)]}} \, \times
\nonumber
\eea
\be
\times \left(\ba{c}
[\sin(\nu_E)J_{-F}(kr) \pm \cos(\nu_E)J_F(kr)]e^{i\chi_\pm}\\[0.2cm]
\sgn(E)[\sin(\nu_E)J_{1-F}(kr) \mp \cos(\nu_E)J_{-1+F}(kr)]\\ \ea \right).
\ee
Here, the energy dependent parameter ($\nu_E$) is expressed through the 
self-adjoint extension parameter ($\Theta$) via relation
\be
\tan(\nu_E)=\sgn(E)\biggl({k\over2\mu}\biggr)^{2F-1}\,
{\Gamma(1-F)\over\Gamma(F)}\tan\bigl(s{\Theta\over2}+{\pi\over4}\bigr),
\ee
$\mu>0$ is the parameter of the dimension of mass,
which is introduced just to scale different irregular behaviours of
the $f$ and $g$ components, and $\Gamma(u)$ is the Euler gamma function; 
note that, since Eqs.(38) and (40) are periodic in
$\Theta$ with period $2\pi$, all permissible values of $\Theta$ can be
restricted, without a loss of generality, to range
$-\pi\leq\Theta\leq\pi$.

Concluding this section, let us note that conditions (38) and (40) coincide 
in the case of
\be
\cos\Theta=0,
\ee
due to vanishing of either their left-hand sides (at $\Theta=s{\pi\over2}$) or  
their right-hand sides (at $\Theta=-s{\pi\over2}$) 
.Thus, 
both parity and chiral symmetry are conserved in the case of Eq.(43).

\section{Chiral symmetry breaking}

Let us consider vacuum polarization effects under the parity invariant 
condition (40). Using the complete set of solutions (35), (36) and (41), we can 
evaluate current ${\mbox{\boldmath $\J$}}(\x)$ (8). Since angular component
\be
\J_{\varphi}(\x)= {i\over4}\,\tr\langle\x|
\gamma_{\varphi}\,|H|^{-1}|\x\rangle
\ee
vanishes, there remains only radial component
\be
\J_{r}(\x)= {i\over4}\,\tr\langle\x|
\gamma_{r}\,|H|^{-1}|\x\rangle,
\ee
which can be obviously presented as a sum over the upper and lower 
inequivalent irreducible representations:
\be
\J_{r}(\x) = \J_{r}^{+}(\x) + \J_{r}^{-}(\x) .
\ee
We find that that the contribution of regular modes (35) and (36) is 
cancelled,
\be
[\J_{r}^{+}(\x)]_\REG = - [\J_{r}^{-}(\x)]_\REG ,
\ee
while the contribution of irregular mode (41) survives,
\be
[\J_{r}^{+}(\x)]_\IRREG = [\J_{r}^{-}(\x)]_\IRREG; 
\ee
consequently, we get
\bea
\J_{r}(\x) = 
-{1\over4\pi}\int\limits_0^\infty dk
\Bigg\{ A\biggl({k\over\mu}\biggr)^{2F-1}
\left[\Lp+\Lm\right]J_{-F}(kr)J_{1-F}(kr)+
\nonumber
\eea
\bea   
+\left[\Lp-\Lm\right][J_F(kr)J_{1-F}(kr)-J_{-F}(kr)J_{-1+F}(kr)]-
\nonumber
\eea
\be
-A^{-1}\biggl({k\over\mu}\biggr)^{1-2F}\left[\Lp+\Lm\right]
J_F(kr)J_{-1+F}(kr)\Bigg\},
\ee
where
\be
A=2^{1-2F}{\Gamma(1-F)\over\Gamma(F)}\tan\left(s{\Theta\over2}+
{\pi\over4}\right) 
\ee
and
\be
L_{(\pm)}=2^{-1}\bigl\{\cos(F\pi)\pm\cosh\bigl[(2F-1)\ln({k\over\mu})+
\ln A\bigr]\bigr\}^{-1};
\ee
incidentally, one can verify that current 
${\mbox{\boldmath $\I$}}(\x)$ (11) vanishes.
Extending the integrand in Eq.(49) to the complex $k$-plane, using the 
Cauchy theorem to deform the contour of integration (for more details see 
Refs.\cite{Sit97,Sit99}) and introducing the 
dimensionless integration variable, we recast Eq.(49) into the form
\be 
\J_{r}(\x) = {\sin(F\pi)\over\pi^3r^2}\int\limits_0^\infty dw\,
{K_F(w)K_{1-F}(w)\over\cosh[(2F-1)\ln({w\over\mu r})+\ln A]},
\ee
where
$$K_\rho(w)={\pi\over2\sin(\rho\pi)}[I_{-\rho}(w)-I_\rho(w)]$$ 
is the Macdonald function of order $\rho$ ($I_\rho(w)$ is the modified 
Bessel function of order $\rho$). Taking into account the invariance of 
Eq.(52) under $s\rightarrow -s$, we rewrite it as
\be
\J_{r}(\x) = {\sin(\fo \Phi^{(0)}\fc\pi)\over \pi^3r^2}\int\limits_0^\infty dw\,
{K_{\fo \,\Phi^{(0)}\,\fc}(w)K_{1-\fo \,\Phi^{(0)}\,\fc}(w)
\over \cosh[(2\fo \Phi^{(0)}\fc-1)\ln \bigl({w\over\mu
r}\bigr)+\ln\bar{A}]},
\ee
where
\be
\bar{A}=2^{1-2\fo \,\Phi^{(0)}\,\fc}\, {\Gamma(1-\fo \Phi^{(0)}\fc)
\over\Gamma(\fo \Phi^{(0)}\fc)} \tan\bigl({\Theta\over2}+{\pi\over4}\bigr).
\ee
                                    
Finally, taking into account Eqs.(7)-(8) and relations
\be
\J_{\varphi}(\x)= 0, \qquad 
{\mbox{\boldmath $\nabla$}}\cdot{\mbox{\boldmath $\J$}}(\x) = 
r^{-1}\partial_{r}[r\J_{r}(\x)],
\ee
we get the following expression for vacuum condensate (5):
\be
\C(\x)=-{\sin(\fo \Phi^{(0)}\fc\pi)\over \pi^3r^2}\int\limits_0^\infty dw\,
w{K^2_{\fo \,\Phi^{(0)}\,\fc}(w)+K^2_{1-\fo \,\Phi^{(0)}\,\fc}(w)
\over \cosh[(2\fo \Phi^{(0)}\fc-1)\ln \bigl({w\over\mu
r}\bigr)+\ln\bar{A}]}.
\ee

Note that the integral in Eq.(53) is ill-defined at half-integer values of the 
vortex flux (at $\fo \Phi^{(0)}\fc ={1\over{2}}$ ). On the contrary, the integral 
in Eq.(56) is well defined at all values of the vortex flux. In particular, 
taking into account relation
\be
\bar{A}|_{\fo \,\Phi^{(0)}\,\fc ={1\over2}}=
\tan\bigl({\Theta\over2}+{\pi\over4}\bigr),
\ee
we get
\be
\C(\x)|_{\fo \,\Phi^{(0)}\,\fc ={1\over2}}=-{\cos\Theta\over 2\pi^2 r^2}.
\ee
Since current ${\mbox{\boldmath $\J$}}(\x)$ plays a merely supplementary role 
and the quantity of physical significance is condensate $\C(\x)$, one can just 
pay no attention to a divergence of the integral in Eq.(53) at
$\fo \Phi^{(0)}\fc ={1\over{2}}$. However, this divergence reveals itself 
when considering the total condensate,
\be
\C= \int d^2x\, \C(\x),
\ee
which can be rewritten as
\be
\C= 2\pi \bigl\{[r\J_{r}(\x)]|_{r=\infty} - [r\J_{r}(\x)]|_{r=0}\bigr\}.
\ee
Therefore, we show in Appendix B, how regularization with the help of 
mass parameter $M$ can be introduced consistently. In particular, after the 
removal of the regularization parameter ($M\rightarrow0$) we find
\be
\sgn(\cos\Theta) r\J_{r}(\x)|_{\fo \,\Phi^{(0)}\,\fc ={1\over2}} = \infty .
\ee
We find also that the regularized version of Eq.(60),
\be
\C^{(M)}=2\pi \bigl\{[r\J_{r}(\x|M)]|_{r=\infty} - [r\J_{r}(\x|M)]|_{r=0}\bigr\}, 
\ee
takes an infinite value at $\fo \Phi^{(0)}\fc ={1\over{2}}$. This 
value is in accordance with the direct evaluation of total condensate (59) 
using the explicit form of Eq.(56), which yields:
\be
\C=-{\sgn(\cos\Theta)\over|2\fo\Phi^{(0)}\fc -1|}.
\ee  
Note that all above quantities are certainly vanishing in the case when 
Eq.(43) holds.

What about other vacuum polarization effects? The vacuum energy density 
can be defined as
\be
\varepsilon^\ren(\x)=-{1\over2} \,\lim_{z\rightarrow -{1\over2}} \,
\lim_{M\rightarrow 0} \,\,
[\zeta_\x(z|M)-\zeta_\x^{(0)}(z|M)] ,
\ee
where
\be
\zeta_\x(z|M)=\tr \langle\x|\,|H^2+M^2|^{-z}|\x\rangle
\ee
is the zeta function density, $M$ is the mass parameter which is 
introduced to regularize the infrared divergence, while the ultraviolet 
divergence is regularized by means of complex parameter $z$: functions
$\zeta_\x(z|M)$ and $\zeta_\x^{(0)}(z|M)$ are evaluated at $\Re z > 1$
and then analytically continued to the whole complex $z$-plane (see Refs.
\cite{Sal,Dow,Hawk}). Note 
that the zeta function density in the trivial background takes form
\be
\zeta_\x^{(0)}(z|M)={|M|^{2(1-z)}\over \pi(z-1)}.
\ee
Let us also define the conventional vacuum current,
\be
\j(\x)=-{1\over2}\,\tr
\langle\x|\gamma^0\,{\mbox{\boldmath $\gamma$}}\,\sgn(H)|\x\rangle,
\ee
in particular, its angular component,
\be
j_\varphi(\x)=-{1\over2}\,\tr
\langle\x|\gamma^0\gamma_\varphi\,\sgn(H)|\x\rangle.
\ee

Similarly to the evaluation of condensate (56), we get
\bea
\zeta_\x(z|M) = {|M|^{2(1-z)}\over\pi(z-1)} +{4\sin(\fo \Phi^{(0)}\fc\pi)
\over\pi^3}\,{\sin(z\pi)\over z-1} r^{2(z-1)}\,\times
\nonumber
\eea
\be
\times \int\limits_{|M|r}^\infty dw\,
(w^2-M^2r^2)^{1-z}K_{\fo \,\Phi^{(0)}\,\fc}(w)K_{1-F}(w)+$$
$$+{2\sin(\fo \Phi^{(0)}\fc\pi)\over\pi^3} \sin(z\pi)r^{2(z-1)} 
\int\limits_{|M|r}^\infty dw\, w(w^2-M^2r^2)^{-z}
\bigl\{ K_{\fo \,\Phi^{(0)}\,\fc}^2(w)+K_{1-\fo \,\Phi^{(0)}\,\fc}^2(w)+$$
$$+\bigl[ K_{\fo \,\Phi^{(0)}\,\fc}^2(w)-K_{1-\fo \,\Phi^{(0)}\,\fc}^2(w)\bigr]
\tanh\bigl[ (2\fo \Phi^{(0)}\fc-1)\ln({w\over\mu r})+\ln \bar{A}\bigr]\bigr\},
\ee
and, consequently,
\bea
\varepsilon^\ren(\x)  =  {\sin(\fo \Phi^{(0)}\fc\pi)\over \pi
r^3} \left\{ {{1\over2}-\fo \Phi^{(0)}\fc\over 6\cos(\fo \Phi^{(0)}\fc\pi)}
\left[{3\over4}-\fo \Phi^{(0)}\fc(1-\fo \Phi^{(0)}\fc)\right]+\right.
\nonumber
\eea
\be + \left. {1\over\pi^2} \int\limits_0^\infty dw\,
w^2\bigl[K^2_{\fo \,\Phi^{(0)}\,\fc}(w) -K^2_{1-\fo \,\Phi^{(0)}\,\fc}(w) \bigr]
\tanh\biggl[(2\fo \Phi^{(0)}\fc-1)\ln\bigl({w\over\mu
r}\bigr)+\ln\bar{A}\biggr]\right\}.
\ee
Also we get 
\bea
j_\varphi(\x)  =  {\sin(\fo \Phi^{(0)}\fc\pi)\over \pi r^2}
\left\{ {(\fo \Phi^{(0)}\fc-{1\over2})^2\over 2\cos(\fo \Phi^{(0)}\fc\pi)}-
\right. 
\nonumber
\eea
\be
-\left.{2\over \pi^2}\int\limits_0^\infty dw\, w
K_{\fo \,\Phi^{(0)}\,\fc}(w)K_{1-\fo \,\Phi^{(0)}\,\fc}(w) \tanh\biggl[
(2\fo \Phi^{(0)}\fc-1)\ln\bigl({w\over\mu
r}\bigr)+\ln\bar{A}\biggr]\right\}. 
\ee
Since the radial component of the vacuum current is not induced, 
Eqs.(56), (70) and (71) comprise all effects of the vacuum polarization in 
background (14)-(15) under condition (40).
At half-integer values of the vortex flux, Eqs.(70) and (71) take form
\be
\varepsilon^\ren(\x)|_{\fo \,\Phi^{(0)}\,\fc ={1\over2}}={1\over12\pi^2r^3}
\ee
and
\be
j_\varphi(\x)|_{\fo \,\Phi^{(0)}\,\fc ={1\over2}}=-{\sin\Theta\over 2\pi^2r^2};
\ee
note the $\Theta$ independence of Eq.(72).

We conclude this section by noting that the vacuum polarization effects
violate translational invariance but remain invariant with respect to a 
rotation around the vortex, depending only on the distance from the vortex. At
large distances they are decreasing by power law:
\bea
\C(\x)
\mathop{=}\limits_{r\rightarrow \infty}
-{\sin(\fo \Phi^{(0)}\fc\pi)\over \pi^2r^2}\,\times
\nonumber
\eea
\be
\times
 \left\{
\ba{cc}
(\mu r)^{2\fo \,\Phi^{(0)}\,\fc-1} \bar{A}^{-1} {\Gamma\bigl({3\over2}-
\fo \,\Phi^{(0)}\,\fc\bigr)\Gamma\bigl({3\over2}-2\fo \,\Phi^{(0)}\,\fc\bigr)\over
\Gamma(1-\fo \,\Phi^{(0)}\,\fc)}, & 0<\fo \Phi^{(0)}\fc<{1\over2}\\[0.4cm]
(\mu r)^{1-2\fo \,\Phi^{(0)}\,\fc}\bar{A} {\Gamma(\fo \,\Phi^{(0)}\,\fc+{1\over2})
\Gamma(2\fo \,\Phi^{(0)}\,\fc-{1\over2})\over \Gamma(\fo \,\Phi^{(0)}\,\fc)},&
{1\over2}<\fo \Phi^{(0)}\fc<1\\ \ea \right. \,  ,
\ee
\bea
\varepsilon^\ren(\x)
\mathop{=}\limits_{r\rightarrow \infty}
{\tan(\fo \Phi^{(0)}\fc\pi)\over2\pi r^3} \biggl(\fo \Phi^{(0)}\fc-{1\over2} \biggr)\, \times
\nonumber
\eea
\be
\times \left[{1\over3}\fo \Phi^{(0)}\fc(1-\fo \Phi^{(0)}\fc)-{1\over4}+{1\over2}|
\fo \Phi^{(0)}\fc-{1\over2}|\right]\,,
\ee
\be
j_\varphi(\x)
\mathop{=}\limits_{r\rightarrow \infty}
{\tan(\fo \Phi^{(0)}\fc\pi)\over2\pi r^2} |\fo \Phi^{(0)}\fc-{1\over2}| \left(|
\fo \Phi^{(0)}\fc-{1\over2}|-1\right).
\ee

\section{Parity breaking}

Let us consider vacuum polarization effects under the chiral invariant 
condition (38). Using the complete set of solutions (35), (36) and (39), we 
find immediately 
\be
\J_{r}^{+}(\x) = - \J_{r}^{-}(\x),
\ee
and, consequently, current
${\mbox{\boldmath $\J$}}(\x)$ (8) vanishes.
Concerning current
${\mbox{\boldmath $\I$}}(\x)$ (11), we find
\be
\I_{\varphi}(\x)= 0
\ee
and
\be
\I_{r}^{+}(\x) = \I_{r}^{-}(\x).
\ee
The contribution of regular solutions (35) and (36) is 
cancelled upon summation over the sign of energy, while 
the contribution of irregular solution (39) survives, resulting in
\be
\I_{r}(\x) = {s\,\sin(\fo \Phi^{(0)}\fc\pi)\over \pi^3r^2}\int\limits_0^\infty dw\,
{K_{\fo \,\Phi^{(0)}\,\fc}(w)K_{1-\fo \,\Phi^{(0)}\,\fc}(w)
\over \cosh[(2\fo \Phi^{(0)}\fc-1)\ln \bigl({w\over\mu
r}\bigr)+\ln\bar{A}]}.
\ee
Consequently, we get the following expression for vacuum condensate (10):
\be
\P(\x)=-{s\,\sin(\fo \Phi^{(0)}\fc\pi)\over \pi^3r^2}\int\limits_0^\infty dw\,
w{K^2_{\fo \,\Phi^{(0)}\,\fc}(w)+K^2_{1-\fo \,\Phi^{(0)}\,\fc}(w)
\over \cosh[(2\fo \Phi^{(0)}\fc-1)\ln \bigl({w\over\mu
r}\bigr)+\ln\bar{A}]}.
\ee
As well as in the case considered in the previous section, energy 
density (70) and current (71) are also induced in the vacuum. However, there are 
additional vacuum polarization effects in the case of the chiral invariant 
condition (38).

Let us define vacuum fermion number density
\be
\N(\x)=-{1\over2}\,\tr\langle\x|\,\sgn(H)|\x\rangle, 
\ee
vacuum spin density
\be
\S(\x) = {1\over8i}\,\tr\langle\x|\,[\gamma^1,\gamma^2]_{-}\,\,\sgn(H)|\x\rangle
\ee
and vacuum angular momentum density
\be 
\M(\x)={i\over2}\,\tr\langle\x|(\x\times\bpar - {1\over4}[\gamma^1,\gamma^2]_{-})
\,\sgn(H)|\x\rangle .
\ee
As it has been already noted in Introduction, the vacuum spin 
density is directly related to the parity breaking vacuum condensate:
\be
\S(\x)={1\over2}\,\P(\x).
\ee
As to remaining vacuum densities (82) and (84), in background (14)-(15) 
under condition (38), they take form 
\be
\N(\x)=-{s\,\sin(\fo \Phi^{(0)}\fc\pi)\over \pi^3r^2}\int\limits_0^\infty dw\,
w{K^2_{\fo \,\Phi^{(0)}\,\fc}(w)-K^2_{1-\fo \,\Phi^{(0)}\,\fc}(w)
\over \cosh[(2\fo \Phi^{(0)}\fc-1)\ln \bigl({w\over\mu
r}\bigr)+\ln\bar{A}]}
\ee
and
\be
\M(\x)=\left(\io\Phi^{(0)}\ic+{1\over2}\right)\,\N(\x) .
\ee
Thus, contrary to the parity breaking condensate , 
\be
\P(\x)|_{\fo \,\Phi^{(0)}\,\fc ={1\over2}}=-{s\,\cos\Theta\over 2\pi^2 r^2},
\ee
the vacuum fermion number and angular momentum vanish at half-integer
values of the vortex flux 
. At large 
distances from the vortex, we get 
\bea
\N(\x)
\mathop{=}\limits_{r\rightarrow \infty}
-\left(\fo \Phi^{(0)}\fc-{1\over2}\right){\sin(\fo \Phi^{(0)}\fc\pi)\over \pi^2r^2}\, \times
\nonumber
\eea
\be
\times
\left\{\ba{cc}
(\mu r)^{2\fo \,\Phi^{(0)}\,\fc-1} \bar{A}^{-1} {\Gamma\bigl({3\over2}-
\fo \,\Phi^{(0)}\,\fc\bigr)\Gamma\bigl({3\over2}-2\fo \,\Phi^{(0)}\,\fc\bigr)\over
\Gamma(2-\fo \,\Phi^{(0)}\,\fc)}, & 0<\fo \Phi^{(0)}\fc<{1\over2}\\[0.4cm]
(\mu r)^{1-2\fo \,\Phi^{(0)}\,\fc}\bar{A} {\Gamma(\fo \,\Phi^{(0)}\,\fc+{1\over2})
\Gamma(2\fo \,\Phi^{(0)}\,\fc-{1\over2})\over \Gamma(1+\fo \,\Phi^{(0)}\,\fc)},&
{1\over2}<\fo \Phi^{(0)}\fc<1\\ \ea \right. \, .
\ee

Finally, defining the global vacuum characteristics, 
\be
\P= \int d^2x\, \P(\x), \, \,\N= \int d^2x\, \N(\x), \,\,\M= \int d^2x\, \M(\x),
\ee
we get
\be
\P=-{s\,\sgn(\cos\Theta)\over|2\fo\Phi^{(0)}\fc -1|} ,
\ee  
\be
\N=-{1\over2} s\,\sgn\left[\left(\fo\Phi^{(0)}\fc -{1\over2}\right)\cos\Theta\right] ,
\ee  
\be
\M=-{1\over2} s \left(\io\Phi^{(0)}\ic+{1\over2}\right)
\sgn\left[ \left(\fo\Phi^{(0)}\fc -{1\over2}\right)\cos\Theta\right].
\ee

\section{Absence of twodimensional anomaly}

Omitting the time dimension (i.e. putting $x^0 = const$), let us define 
twodimensional Euclidean effective action functional
\be
S^{\rm eff}[\V(\x)]=-\ln \int d\Psi \,d\Psi^\dagger\,
\exp(-\int d^2 x \,\Psi^\dagger H \Psi ) = - \ln \det H  .
\ee
The last equation is formally invariant 
under transformation
\be
\Psi \rightarrow e^{i\omega\Gamma} \Psi, \qquad 
\Psi^\dagger \rightarrow \Psi^\dagger e^{i\omega\Gamma},
\ee
where $\Gamma$ is a matrix which anticommutes with the Hamiltonian,
\be
[H,\Gamma]_+=0 ,\qquad  \tr\Gamma=0 ,\qquad  \Gamma^2=I .
\ee
The invariance under the localized (coordinate dependent) generalization 
of transformation (95) corresponds to the conservation law:
\be
{\mbox{\boldmath $\nabla$}}\cdot{\bf J}^3(\x) = 0,
\ee
where
\be
{\bf J}^3(\x) = i\,\tr\langle\x|\,\gamma^0
{\mbox{\boldmath $\gamma$}}\,\Gamma\,H^{-1}|\x\rangle.
\ee
However, both functional (94) and current (98) are ill-defined, suffering from 
ultraviolet as well as infrared divergences. Performing the regularization 
of divergences with the use of the zeta function method \cite{Sal,Dow,Hawk}, 
one arrives at, instead of Eq.(97), the following relation:
\be
{\mbox{\boldmath $\nabla$}}\cdot{\bf J}^3(\x) = 2 \,\lim_{z\rightarrow 0} \,
\lim_{M\rightarrow 0} \,\,
\tilde{\zeta}_\x(z|M),
\ee
where
\be
\tilde{\zeta}_\x(z|M)=\tr\langle\x|\Gamma\,(H^2+M^2)^{-z}|\x\rangle
\ee
is the modified zeta function density (compare with Eq.(65)).

In the reducible $4\times 4$ representation of the Clifford algebra 
(see Eq.(21)), the role of $\Gamma$ can be played by each of the 
following four matrices:
\beq
\gamma^0, \qquad \gamma^0\gamma^3, \qquad \gamma^0\gamma^5, \qquad  
{1\over2}\gamma^0[\gamma^3,\gamma^5]_{-} \,.
\eeq
However, only one choice,
\be
\Gamma = {1\over2}\gamma^0[\gamma^3,\gamma^5]_{-},
\ee
can lead to the nonvanishing modified zeta function density and, thus, to the 
anomaly in the conservation of the appropriate current,
\be
{\bf J}^3(\x) = {1\over2i}\,\tr\langle\x|\,
{\mbox{\boldmath $\gamma$}}\,[\gamma^3,\gamma^5]_{-}\,H^{-1}|\x\rangle.
\ee
Reflecting a generic relationship between Eq.(102) and current
${\mbox{\boldmath $\I$}}(\x)$ (11), this anomaly is usually called as the 
parity anomaly.

The question that we would like to address in the present section is: whether
singular magnetic vortex (14)-(15) induces the parity anomaly or not? It is 
clear that the parity anomaly is absent under the parity invariant condition 
(40). So there remains to check the absence of the parity anomaly under the 
chiral invariant condition (38). Similarly to that in the above sections, we 
find that
\be
\tilde{\zeta}_\x(z|M)={1\over2} \, \tr\langle\x|\gamma^0[\gamma^3,\gamma^5]_{-}\,
(H^2+M^2)^{-z}|\x\rangle
\ee
under condition (38) takes form
\bea
\tilde{\zeta}_\x(z|M) = {2\,s\,\sin(\fo \Phi^{(0)}\fc\pi)\over\pi^3} \sin(z\pi)r^{2(z-1)}\, \times
\nonumber
\eea
\be
\times
\int\limits_{|M|r}^\infty dw\, w(w^2-M^2r^2)^{-z}
\bigl\{ K_{\fo \,\Phi^{(0)}\,\fc}^2(w)-K_{1-\fo \,\Phi^{(0)}\,\fc}^2(w)+$$
$$+\bigl[ K_{\fo \,\Phi^{(0)}\,\fc}^2(w)+K_{1-\fo \,\Phi^{(0)}\,\fc}^2(w)\bigr]
\tanh\bigl[ (2\fo \Phi^{(0)}\fc-1)\ln({w\over\mu r})+\ln \bar{A}\bigr]\bigr\}.
\ee
Hence, one obtains immediately
\be
\lim_{z\rightarrow 0} \,\,\tilde{\zeta}_\x(z|M) = 0  \qquad  (M \neq 0)  .
\ee
It is more instructive to take limit $M\rightarrow 0$ first and then to 
consider limit $z\rightarrow 0$. Thus, we get
\be
\tilde{\zeta}_\x(z|0)={s\,\sin(\fo \Phi^{(0)}\fc\pi)\over\pi^2}
\, r^{2(z-1)}\Bigg\{ {\sqrt{\pi}(\fo \Phi^{(0)}\fc-{1\over2})\over 
\Gamma(z)\Gamma({3\over2}-z)} \,\Gamma(\fo \Phi^{(0)}\fc-z)
\Gamma(1-\fo \Phi^{(0)}\fc-z)+$$
$$ +{2\sin(z\pi)\over\pi}\,\int\limits_0^\infty dw\, w^{1-2z}\bigl[K^2_{\fo \,\Phi^{(0)}\,\fc}(w) 
+K^2_{1-\fo \,\Phi^{(0)}\,\fc}(w) \bigr]
\tanh\biggl[(2\fo \Phi^{(0)}\fc-1)\ln\bigl({w\over\mu
r}\bigr)+\ln\bar{A}\biggr] \Bigg\} ;
\ee
in particular, at half-integer values of the vortex flux:
\be
\tilde{\zeta}_\x(z|0)\big|_{\fo \,\Phi^{(0)}\,\fc ={1\over2}}={s\,\sin\Theta\over
\pi^{\sss3\over\sss2}}\, {\Gamma({1\over2}-z)\over\Gamma(z)} r^{2(z-1)};
\ee
and in the case when Eq.(43) holds:
\bea
\tilde{\zeta}_\x(z|0)=\pm{s\,\sin(\fo \Phi^{(0)}\fc\pi)\over\pi^{\sss3\over\sss2}}\,\times
\nonumber
\eea
\be
\times
{\Gamma({3\over2}-z\pm \fo \Phi^{(0)}\fc\mp{1\over2})\Gamma({1\over2}-z\mp
\fo \Phi^{(0)}\fc\pm{1\over2})\over \Gamma(z)\Gamma({3\over2}-z)}\, r^{2(z-1)} , \quad
\Theta=\pm s{\pi\over2}; 
\ee
Consequently, we obtain
\be
\tilde{\zeta}_\x(0|0)=0, \qquad \x\neq0, 
\ee
which ensures the validity of Eq.(97) everywhere on the plane with the 
puncture at $\x=0$.

\section{Conclusion}

Before discussing our results, let us recall the well-known fact: a
regular configuration of external magnetic field strength $\bpar\times\V(\x)$ 
in 2+1-dimensional space-time induces the parity anomaly \cite{Schw,Shei,Jack},
\be
{\mbox{\boldmath $\nabla$}}\cdot{\bf J}^3(\x) = 
{2\,s\over\pi}\, \bpar\times\V(\x) \, ,
\ee
as well as the chiral symmetry breaking vacuum condensate \cite{Gus},
\be                             
\C(\x)=-{1\over2\pi}\, |\bpar\times\V(\x)| \, .
\ee
These relations exhibit a direct (or local) impact of an external field strength 
on a quantized fermion field. If one excludes the spatial region of nonvanishing 
field strength and imposes a physically sensible condition at the boundary 
of the excluded region, what happens then with a quantized fermion field 
in the remaining part of space? Basing merely on Eqs.(110) and (111), one could 
expect that both the anomaly and the condensate vanish in the region of 
vanishing field strength. However, as it is shown in the present paper, these 
naive expectations are justified for the anomaly only (see Section 5), whereas 
the condensate and other vacuum quantum numbers appear to be nonvanishing, thus 
exhibiting an indirect (or nonlocal) impact of an external field strength, 
which may be regarded (see, e.g., Ref.\cite{Gro}) as a leak through the 
boundary of the excluded region. 

To be more precise, we consider the situation when the volume of the excluded 
region is shrunk to zero, while the global characteristics of an external field 
strength (flux) is retained nonvanishing. Therefore, singular magnetic vortex 
(14)-(15) is taken as an external field configuration. The boundary condition 
at the location of the vortex has to ensure self-adjointness of the Hamiltonian. 
We find two sets of boundary conditions -- one is chiral invariant (38) and 
another is parity invariant (40); each set is labelled by the self-adjoint 
extension parameter. Thus, vacuum polarization effects in the background of a 
singular magnetic vortex are depending both on the vortex flux and the 
self-adjoint extension parameter.

As it should be expected, the vacuum polarization effects remain invariant 
under the transition to the equivalent representation of the Clifford algebra 
(i.e. do not depend on $\chi_{\pm}$). If the parity invariant condition (40) is 
imposed, then the vacuum polarization effects are invariant under the 
transition to the inequivalent representation of the Clifford algebra (i.e. 
do not depend on $s$). They comprise chiral symmetry breaking vacuum condensate 
(56), vacuum energy density (70) and vacuum current (71). If the chiral invariant 
condition (38) is imposed, then the vacuum polarization effects are either 
invariant or changing sign under $s\rightarrow -s$. The invariant effects 
comprise vacuum energy density (70) and vacuum current (71), while the 
changing sign effects comprise parity breaking vacuum condensate (81), vacuum 
fermion number density (86) and vacuum angular momentum density (87). All 
vacuum polarization effects are decreasing as inverse powers (with
integer exponents in the cases of vacuum energy density and current, and with
fractional exponents otherwise) at large distances 
from the vortex, see Eqs.(74)-(76) and (89).
Total vacuum condensates are finite at non-half-integer values of the vortex 
flux, see Eqs.(63) and (91). 
Total vacuum fermion number and angular momentum are finite, vanishing at
half-integer values of the vortex flux, see Eqs.(92) and (93). 

It should be noted that both conditions (38) and (40) become scale invariant
at half-integer values of the vortex flux. Thus, Eq.(58) exhibits a scale 
invariant pattern of chiral symmetry breaking. Other scale invariant effects 
are given by vacuum energy density (72), vacuum current (73) and  parity 
breaking vacuum condensate (88). The infinite value of total vacuum condensates 
(63) and (91) at $\fo \Phi^{(0)}\fc ={1\over{2}}$ is a consequence of scale 
invariance.

Finally, let us discuss the case of Eq.(43) when conditions (38) and 
(40) coincide. A distinctive feature in this case is that two of the four
components of wave function (31) become regular for all $n$: if
$\Theta=s\,{\pi\over2}$, then the $g_n^\pm$ components are regular, and,
if $\Theta=-s\,{\pi\over2}$, then the $f_n^\pm$ components are regular.
Parity and chiral symmetry, as well as scale symmetry, are conserved, and 
only energy density and current are induced in the vacuum:
\bea
\varepsilon^\ren(\x) & = & {\tan(\fo\Phi\fc \pi) \over{2}\pi
r^3} \left(\fo\Phi\fc -{1\over2}\right) \left[{1\over 3}\fo\Phi\fc
\left(1-\fo\Phi\fc \right)-{1\over 4}\mp {1\over 2}\left(\fo\Phi\fc
-{1\over 2}\right) \right], \nonumber\\[0.3cm]
&&\Theta=\pm{\pi\over{2}},  \\[0.3cm]
j_\varphi(\x) & = & {\tan(\fo\Phi\fc \pi)\over 2\pi r^2}
\left( \fo\Phi\fc -{1\over2}\right) \left(\fo\Phi\fc
-{1\over2}\pm1\right), \quad \Theta=\pm{\pi\over2}. 
\eea

\section*{Acknowledgements}

I am grateful to H.~Leutwyler, V.A.~Miransky and W.~Thirring for stimulating
discussions and interesting remarks. The research was supported by the
State Foundation for Fundamental Research of Ukraine (project 2.4/320).

\section*{Appendix A}
\def\theequation{A.\arabic{equation}}
\setcounter{equation}{0}

The partial Hamiltonian corresponding to $n = n_0$ takes form
\be
h=\left(\ba{cc}
h_+& 0\\
0& h_-\\ \ea \right),
\ee
where (see Eq.(32))
\be
h_\pm=\left(\begin{array}{cc}0&e^{i\chi_\pm}[\partial_r+(1-F)r^{-1}]\\
e^{-i\chi_\pm}(-\partial_r-Fr^{-1})&0\end{array}\right).
\ee

Let $h$ be the operator in the form of Eqs.(A.1)-(A.2), which acts on
the domain of functions $\xi^{(0)}(r)$ that are regular at $r=0$. Then its 
adjoint $h^{\dagger}$ which is defined by relation
\be
\int^\infty_0 dr\,r[h^\dagger\xi(r)]^\dagger \xi^{(0)}(r) =
\int^\infty_0 dr\,r[\xi(r)]^\dagger[h\xi^{(0)}(r)]
\ee
acts on the domain of functions $\xi(r)$ that are not necessarily
regular at $r=0$. So the question is, whether the domain of definition
of $h$ can be extended, resulting in both the operator and its
adjoint being defined on the same domain of functions? To answer this,
one has to construct the eigenspaces of $h^\dagger$
with complex eigenvalues. They are spanned by the linearly independent
square-integrable solutions correspoding to the pair of purely imaginary
eigenvalues,
\be
h^\dagger \xi^{(1)}(r) = i\mu\xi^{(1)}(r),  \qquad
h^\dagger \xi^{(2)}(r) = - i\mu\xi^{(2)}(r),
\ee
where $\mu>0$ is inserted merely for the dimension reasons. It can be shown 
that, in the case of Eqs.(A.1)-(A.2), only one pair of such solutions exists:
\be
\xi^{(1)}(r) =
\left(\ba{c}
\xi^{(1)}_+(r)\\
\xi^{(1)}_-(r)\\ \ea
\right),   \qquad 
\xi^{(2)}(r) =
\left(\ba{c}
\xi^{(2)}_+(r)\\
\xi^{(2)}_-(r)\\ \ea
\right),
\ee
where
\be
\xi^{(1)}_\pm (r) ={1\over N}
\left(\ba{c}
e^{i\chi_\pm}e^{i{\pi\over4}} K_F(\mu r)\\[0.2cm]
e^{-i{\pi\over4}}K_{1-F}(\mu r)\\ \ea \right), \qquad 
\xi^{(2)}_\pm (r) ={1\over N}
\left(\ba{c}
e^{i\chi_\pm}e^{-i{\pi\over4}} K_F(\mu r)\\[0.2cm]
e^{i{\pi\over4}}K_{1-F}(\mu r)\\ \ea \right), 
\ee
$N$ is a certain normalization factor. Thus the deficiency
index of each of operators $h_+$ and $h_-$ is equal to (1,1). Then, according 
to the Weyl - von Neumann theory of self-adjoint operators (see 
Refs.\cite{Akhie,Alb}), the self-adjoint extension of operator $h_\pm$ 
is defined on the domain of functions of the following form
\be
\left(\ba{c}
f_{n_0}^\pm\\
g_{n_0}^\pm\\ \ea
\right) = \xi_\pm^{(0)} + c (\xi_\pm^{(1)} + e^{i\theta_\pm (s)}\xi_\pm^{(2)}),
\ee
where $c$ is a complex constant and $\theta_\pm (s)$ is a real continuous
parameter which depends on $s$. Using the asymptotics of the Macdonald function 
at small values of the variable, we find that operator (A.1)-(A.2) is 
self-adjoint when defined on the domain of functions with the following 
asymptotic behaviour:
\be
\left(\ba{c}
f_{n_0}^+\\
g_{n_0}^+\\
f_{n_0}^-\\
g_{n_0}^-\\ \ea
\right)
\mathop{\sim}_{r\to0}
\left(\ba{c}
e^{i\chi_+}\,\sin[{1\over2}\theta_+(s)+{\pi\over4}]\,2^F\,\Gamma (F)(\mu r)^{-F}\\
\cos[{1\over2}\theta_+(s)+{\pi\over4}]\,2^{1-F}\,\Gamma (1-F)(\mu r)^{-1+F}\\                                  
e^{i\chi_-}\,\sin[{1\over2}\theta_-(s)+{\pi\over4}]\,2^F\,\Gamma (F)(\mu r)^{-F}\\
\cos[{1\over2}\theta_-(s)+{\pi\over4}]\,2^{1-F}\,\Gamma (1-F)(\mu r)^{-1+F}\\ \ea
\right)   .
\ee

If one chooses
\be
\theta_+(s) = \theta_-(s) = s\theta ,
\ee
then asymptotics (A.8) is invariant under the chiral symmetry transformation 
(25) and the following relation holds
\be
\cos\bigl(s{\theta\over2}+{\pi\over4}\bigr) \,2^{1-F}\,\Gamma (1-F)\,
\lim_{r\rightarrow 0}(\mu r)^Ff^\pm_{n_0} = 
e^{i\chi_{\pm}}\sin\bigl(s{\theta\over2}+{\pi\over4}\bigr)
\,2^F\,\Gamma (F)\,\lim_{r\rightarrow 0}(\mu r)^{1-F}g^\pm_{n_0}. 
\ee
If one chooses
\be
\theta_+(s) = s\theta ,  \qquad   \theta_-(s) = - s\theta + \pi ,
\ee
then asymptotics (A.8) is invariant under the parity transformation (24) 
and the following relation holds
\be
\cos\bigl(s{\theta\over2}+{\pi\over4}\bigr) \,2^{1-F}\,\Gamma (1-F)\, 
\lim_{r\rightarrow 0}(\mu r)^Ff^\pm_{n_0} = 
\pm e^{i\chi_{\pm}}\sin\bigl(s{\theta\over2}+{\pi\over4}\bigr)
\,2^F\,\Gamma (F)\,\lim_{r\rightarrow 0}(\mu r)^{1-F}g^\pm_{n_0}. 
\ee
Defining new parameter $\Theta$ by means of relation
\be
\tan\bigl(s{\Theta\over2}+{\pi\over4}\bigr) = -2^{2F-1}\,
{\Gamma(F)\over\Gamma(1-F)}\tan\bigl(s{\theta\over2}+{\pi\over4}\bigr),
\ee
we rewrite Eqs.(A.10) and (A.12) in the form of Eqs.(38) and (40) respectively.

\section*{Appendix B}
\def\theequation{B.\arabic{equation}}
\setcounter{equation}{0}

Inserting regulator mass $M$ into Eq.(2), we get
\be
\langle\vac| T\Psi(x)\bar{\Psi}(y)|\vac\rangle = -i\langle x|
(-i\gamma^\mu\nabla_\mu + M)^{-1} |y \rangle .
\ee
Then the regularized condensate takes form
\bea
\C(\x|M)=-{1\over2} \tr\langle\x|\gamma^0\,\sgn(H+\gamma^0 M)|\x\rangle =
\nonumber
\eea
\be
= {i\over4}\,{\mbox{\boldmath $\nabla$}}\,\cdot\tr\langle\x|
{\mbox{\boldmath $\gamma$}}\,(H^2 + M^2)^{-{1\over2}}|\x\rangle -
{1\over2} M \,\zeta_{\x}({1\over2}|M),
\ee
where zeta function density $\zeta_{\x}(z|M)$ is defined by Eq.(65). Defining 
regularized current
\be
{\mbox{\boldmath $\J$}}(\x|M)= {i\over4}\,\tr\langle\x|
{\mbox{\boldmath $\gamma$}}\,(H^2 + M^2)^{-{1\over2}}|\x\rangle,
\ee
we get
\bea
\J_{r}(\x|M) = {\sin(\fo \Phi^{(0)}\fc\pi)\over \pi^3r^2}
\int\limits_{|M|r}^\infty dw{w\over \sqrt{w^2 - M^2 r^2}}\times
\nonumber
\eea
\be
\times {K_{\fo \,\Phi^{(0)}\,\fc}(w)K_{1-\fo \,\Phi^{(0)}\,\fc}(w)
\over \cosh[(2\fo \Phi^{(0)}\fc-1)\ln \bigl({w\over\mu
r}\bigr)+\ln\bar{A}]},
\ee
and
\be
\J_{\varphi}(\x|M) = 0 ;
\ee
consequently,

\bea
{\mbox{\boldmath $\nabla$}}\cdot{\mbox{\boldmath $\J$}}(\x|M) =
-{\sin(\fo \Phi^{(0)}\fc\pi)\over \pi^3r^2}\int\limits_{|M|r}^\infty dw
{w^2\over \sqrt{w^2 - M^2 r^2}}\times
\nonumber
\eea
\be
\times{K^2_{\fo \,\Phi^{(0)}\,\fc}(w)+K^2_{1-\fo \,\Phi^{(0)}\,\fc}(w)
\over \cosh[(2\fo \Phi^{(0)}\fc-1)\ln \bigl({w\over\mu
r}\bigr)+\ln\bar{A}]}.
\ee
In particular, at half-integer values of the vortex flux 
(at $\fo \Phi^{(0)}\fc = {1\over2}$) we get
\be
\J_{r}(\x|M)|_{\fo \,\Phi^{(0)}\,\fc ={1\over2}}={\cos\Theta\over 2\pi^2 r}
K_0 (2|M|r)
\ee
and
\be
{\mbox{\boldmath $\nabla$}}\cdot{\mbox{\boldmath $\J$}}
(\x|M)|_{\fo \,\Phi^{(0)}\,\fc ={1\over2}} =
-{\cos\Theta\over \pi^2 r} |M| K_1 (2|M|r)  ;
\ee
consequently,
\be
\J_{r}(\x|M\rightarrow0)|_{\fo \,\Phi^{(0)}\,\fc ={1\over2}}=
-{\cos\Theta\over 2\pi^2 r}\ln (2|M|r)
\ee
and
\be
{\mbox{\boldmath $\nabla$}}\cdot{\mbox{\boldmath $\J$}}
(\x|M\rightarrow0)|_{\fo \,\Phi^{(0)}\,\fc ={1\over2}} =
-{\cos\Theta\over 2\pi^2 r^2}  .
\ee

Using Eq.(69), we get relation
\bea
\zeta_\x({1\over2}|M)=-{2\over\pi}|M| - {8\sin(\fo \Phi^{(0)}\fc \pi)
\over\pi^3 r} \int\limits_{|M|r}^\infty dw \sqrt{w^2 - M^2 r^2} \, 
K_{\fo \,\Phi^{(0)}\,\fc}(w)K_{1-\fo \,\Phi^{(0)}\,\fc}(w)+
\nonumber
\eea
\bea
+ {2\sin(\fo \Phi^{(0)}\fc\pi)\over\pi^3 r} \int\limits_{|M|r}^\infty
dw {w \over \sqrt{w^2 - M^2 r^2}} \, \bigl\{ K_{\fo \,\Phi^{(0)}\,\fc}^2(w) + 
K_{1-\fo \,\Phi^{(0)}\,\fc}^2(w) +
\nonumber
\eea
\be
+\bigl[ K_{\fo \,\Phi^{(0)}\,\fc}^2(w) - K_{1-\fo \,\Phi^{(0)}\,\fc}^2(w)\bigr]
\tanh\bigl[ (2\fo \Phi^{(0)}\fc-1)\ln({w\over\mu r})+\ln \bar{A}\bigr]\bigr\}, 
\ee
and, consequently,
\be
\lim_{M\rightarrow 0} M \,\zeta_{\x} ({1\over2}|M) = 0.
\ee

Thus, Eqs.(B.4) and (B.6) in the limit of $M\rightarrow 0$ yield Eqs.(53)
and (56), respectively, and Eq.(B.10) coincides with Eq.(58). Note also 
relations
\be
\lim_{r\rightarrow \infty} r\J_{r}(\x|M)|_{\fo \,\Phi^{(0)}\,\fc = {1\over2}} 
= 0
\qquad   (M \neq 0)
\ee
and
\be
\sgn(\cos\Theta)\lim_{r\rightarrow 0} 
r\J_{r}(\x|M)|_{\fo \,\Phi^{(0)}\,\fc ={1\over2}}= \infty 
\qquad   (\Theta \neq \pm{\pi\over2}) ;
\ee
thus, the total condensate at half-integer values of the vortex flux is 
infinite even at finite values of $M$.

\end{document}